\documentclass[preprint,12pt]{elsarticle}

\usepackage{amssymb}
\usepackage{multicol}
\usepackage{booktabs}
\usepackage{graphicx}
\usepackage{tabularx}
\usepackage{geometry}
\usepackage{pdfpages}
\usepackage{subfigure}
\usepackage{amsmath}

\journal{International Journal of Rail Transportation}

\begin{document}

\begin{frontmatter}



\title{Design of railway transition zones: a novel energy-based criterion}


\author{A. Jain, A.V. Metrikine, M.J.M.M. Steenbergen and K.N. van Dalen}

\affiliation{organization={Faculty of Civil Engineering and Geosciences (CEG), Department of Engineering Structures, TU Delft},
            addressline={Stevinweg 1}, 
            city={Delft},
            postcode={2628 CN}, 
            country={The Netherlands}}

\begin{abstract}
Railway transition zones (RTZs) experience higher rates of degradation compared to open tracks, which leads to increased maintenance costs and reduced availability. Despite existing literature on railway track assessment and maintenance, effective design solutions for RTZs are still limited. Therefore, a robust design criterion is required to develop effective solutions. This paper presents a two-step approach for formulation of a design criterion to delay the onset of processes leading to uneven track geometry due to operation driven permanent deformations in RTZs. Firstly, a systematic analysis of each track component in a RTZ is performed by examining spatial and temporal variations in kinematic responses, stresses and energies. Secondly, the study proposes an energy-based criterion to be assessed using a model with linear elastic material behavior, and states that an amplification in the total train energy in the proximity of the transition interface is an indicator of increased (and thus non-uniform) degradation in RTZs compared to the open tracks. The correlation between the total strain energy (assessed in the model with linear material behaviour) and the permanent irreversible deformations is demonstrated using a model with non-linear elastoplastic material behavior of the ballast layer. In the end, it is claimed that minimising the magnitude of total strain energy will lead to reduced degradation and a uniform distribution of total strain energy in each trackbed layer along the longitudinal direction of the track will ensure uniformity in the track geometry.
\end{abstract}




\end{frontmatter}


\section{Introduction}
Railway tracks are subject to constant degradation and require frequent maintenance which in turn leads to high maintenance costs and reduced frequency of moving trains. Consequently, a large proportion (40-75\%) of railway operating costs are spent on track maintenance \cite{1} to ensure passenger comfort and safety. There is plenty of literature focused on assessment and maintenance of railway tracks. However, the problem at hand still remains and is even intensified in countries with soft soil \cite{2}. The frequency and costs of maintenance in railway transition zones (RTZs) is even higher (4-8 times) compared to the open tracks \cite{1,2}. RTZs are areas where the railway track crosses a structure related to a different transportation modality (bridge, road, culvert, etc.) or where the rail experiences major changes in the type of track support structure. Although there have been several studies \cite{3,4,5,39,41} pointing out the cause of increased degradation in these zones as mainly abrupt change in stiffness and differential settlement, there still is a lack of understanding to design an effective solution to this problem. The current approach to deal with the excessive degradation involves some proactive and reactive measures \cite{6,7,8,9,39} that have proved either not as efficient or even counterproductive in some cases. A robust design solution for RTZs is lacking in literature. RTZs are built similarly to regular railway tracks with some modifications addressing the stiffness jump and differential settlement but without a complete knowledge of the variation of dynamic response of each track component. In order to delay the onset of processes leading to uneven track geometry due to non-recoverable permanent deformations in RTZs, there is a need for an effective design solution. The formulation of an effective design solution demands mainly two steps as discussed below.\\

Firstly, there is a need to perform a systematic analysis of each track component in RTZs which involves a detailed study of spatial and temporal variation of kinematic responses, stresses and energies. A railway track is composed of several components and each component serves a specific function in the system. The response of each track component experiences a variation in vertical (depth of the track), transversal (along the width of the track), and longitudinal (along the length of the track) directions. These variations affect the performance of the track components in the track which in turn drives non-uniform degradation and the failure process. In addition to this, the degradation or failure of one component has an effect on the performance of the other components in the track system. Therefore, the damage prediction requires a detailed and systematic study of the behavior of each track component and an in depth understanding of interactions between components. The current literature \cite{8,10,11,12,13,14,15,16,17,37,42}  focuses either on any particular track component (mostly rail, sleepers or ballast) or on one particular track response (mainly vertical displacements, accelerations or stress in ballast layer). However, a detailed analysis of the spatial and temporal variation of the response of each track component in RTZs is lacking in literature. Furthermore, an energy analysis of the track components in RTZs, which could help identify signs of degradation, has not yet been presented in the existing literature. The significance of energy variation will be discussed in the following paragraph, and this paper will investigate it in detail.\\

Secondly, identification of an appropriate design criterion affecting the response of RTZs is essential to design an effective design solution. The current literature evaluates the performance of a RTZ mainly by assessing either permanent vertical track deformations or stresses at the sleeper-ballast interface. However, the question can be asked whether the deformations or the stresses at this particular interface are sufficient to describe the onset of degradation in a RTZ. What precisely leads to faster degradation of these zones compared to the open tracks? The source, spacial extent and location of the degradation in these zones is still unknown. As indicated above, the literature lacks an insight into the spatial and temporal distribution of kinetic and potential energies in railway track components, while this is an indispensable information to answer the question posed. Studies conducted by different authors \cite{18,19,21} have pointed out that both elastic and inelastic strain energy can be associated with material degradation and may aid in identifying potential failure mechanisms. In \cite{18} authors have studied shakedown of soil in different test setups in terms of elastic and plastic strain energies. In \cite{20} authors assessed the susceptibility of railway tracks to degradation by quantifying the mechanical energy dissipated under a moving train axle, but, due to its dimensional constraints, the model could neither demonstrate the variations of these energies in each layer of the substructure nor was focused on railway transition zones. Hence, the current work will investigate the mechanical energy distribution in space and time for RTZs using a model with linear elastic materials and demonstrate a correlation of the predicted responses with irreversible permanent deformation leading to uneven track geometry. In the end, this work will propose a design criterion to minimize the degradation in RTZs.\\

This work is mainly divided in two parts addressing the points mentioned above. In the first part of this work, a 2-dimensional (2-D) model of an embankment bridge transition with linear elastic materials is used to study the spatial and temporal distribution of the kinematic responses, stresses and energies for various track components. As the intent of this paper is to identify a possible design criterion for transition zones, the focus is not so much on the absolute values of the responses under study but their variation in space and time. For the same reason, only one train axle is considered. In the second part of this work, a design criterion is proposed based on the comprehensive analysis presented in the first part. Moreover, a correlation between permanent irreversible deformations and the proposed design criterion is demonstrated using a non-linear elastoplastic material for the ballast layer. It is to be noted that this paper is focused on the investigation of operation driven permanent deformations (excludes autonomous settlements) as the aim is to the study the effects of dynamic amplifications and non-uniformity in the response that may trigger degradation.\\

\section{Methods}
In this paper, a embankment-bridge transition is studied which consists of ballasted track and a concrete bridge. The ballasted track (“soft side”) consists of track components namely rail, rail-pads, sleepers, ballast, embankment and sub-grade underneath. The ballast-less track (“stiff side”) consists of rails connected to sleepers (with under sleeper pads) resting on the concrete bridge. 
\subsection{Geometric model }
\begin{figure}
  \centering  \includegraphics[angle=90,width=\textwidth]{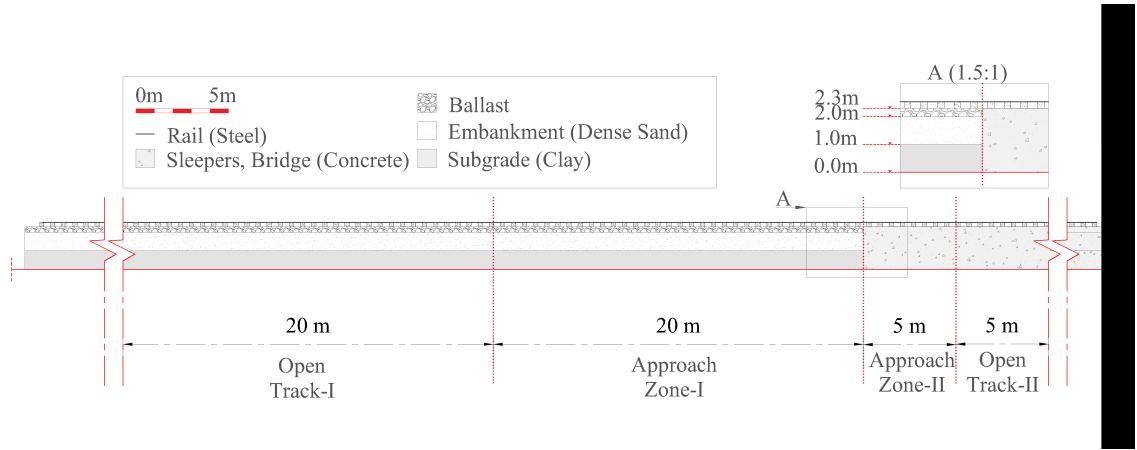}
  \caption{Cross-sectional details of the embankment-bridge transition and the division of zones (OT-I, AZ-I, AZ-II, OT-II) under study.}
  \label{fig1}
\end{figure}

The geometric model is 80 m (132 sleepers) long which consists of 60 m of ballasted track and 20 m of ballast-less track (concrete bridge). For this study, the geometrical model is divided into four zones (see Table-\ref{table1}) with additional 20 m at the beginning and 10 m at the end in order to eliminate the influence of boundaries (left and right extremes) on the results for the reasons discussed in section \ref{NumModel}. Figure \ref{fig1} shows the approach zones (AZ) on both the sides that together constitute the transition zone (TZ), and open track (OT) on both sides are the zones that are free from any transition effects.

\begin{table}[htbp]
\centering
\caption{Details of the zones under study}
\begin{tabularx}{\textwidth}{@{}XXX@{}}
\toprule
Zones & Length [m] & Description \\
\midrule
OT-I  & 20   & open track-soft side     \\
AZ-I  & 20   & approach zone-soft side  \\
AZ-II & 5    & approach zone-stiff side \\
OT-II & 5    & open track-stiff side    \\
\bottomrule
\label{table1}
\end{tabularx}

\end{table}

The geometric model mainly consists of following track components (Figure \ref{fig1}):

\begin{itemize}
\item	Rail: rail profile 54E1 (UIC54) manufactured according European Standard EN 13674-1
\item	Rail-pads: connecting rails to sleepers
\item	Sleepers: 240 mm x 240 mm, concrete
\item	Ballast: 0.3 m deep layer of ballast 
\item	Embankment: 1 m deep dense sand under the ballast 
\item	Natural terrain or subgrade: clayey soil of 1 m depth 
\item	Under Sleeper pads (USPs) \cite{38}: under sleeper pads of 0.01 m thickness under the sleepers on the stiff side
\item	Bridge: concrete bridge with fixed bottom of length 20 m

\end{itemize}

The sleeper spacing adopted is 0.6 m and the first sleeper next to the transition is located at 0.3 m from the interface of ballasted and ballastless track.

\subsection{Numerical Model} \label{NumModel}
A 2-D Finite Element (FE) model was created using ABAQUS, with geometrical details mentioned in the previous section. The railway track components and the bridge were modelled using four-node plane-strain elements and the rail was modelled as a beam. Rayleigh damping for materials (see Table-\ref{table2}) was used to incorporate realistic energy dissipation in the system. The length and depth of the model were chosen such that there is no influence of the wave reflections from the boundaries (extreme right, left and bottom of the system) on results of the zones under study while restricting the vertical displacements to maintain a reasonable value as per literature \cite{22,23}. The depth of the layers under the sleepers was sufficient to reduce dynamic stresses at the bottom of the subgrade to less than 3\% of their values at bottom of the sleepers. The following subsections describe the details regarding the numerical model used for FE analysis in terms of mesh properties, mechanical properties of materials, interactions between the track components, loads, constraints, boundary conditions and analysis procedures.

\subsubsection{Mesh}
The sleeper, ballast, embankment, subgrade and the bridge were discretized \cite{24} using linear quadrilateral elements of type CPE4R (12424 elements) and rail using two-node linear line elements of type B21 (7860 elements) to form a very regular mesh. The most critical zone (AZ-I) under study was meshed finer than rest of the model in order to obtain accurate results. 

\subsubsection{Mechanical properties}
The materials used for all track components are characterised by elastic properties (Young’s modulus, Poisson’s ratio), densities and Rayleigh damping factors \cite{25,26,27,28}. A static analysis was performed in order to tune the elasticity (Young’s modulus) of the USP on the stiff side such that the static vertical displacements on soft and stiff sides are the same. The details of these parameters are mentioned in Table-\ref{table2}.

\begin{table}[htbp]
\caption{Mechanical properties of the track components}
\centering
    \begin{tabularx}{\textwidth}{@{}p{4cm}XXXX>{\centering\arraybackslash}X>{\centering\arraybackslash}X@{}}
    \toprule
    Material & Elasticity Modulus& Density & Poisson’s Ratio & \multicolumn{2}{c}{Rayleigh damping} \\ 
    & \textit{E} {[}MN/$\text{m}^2${]} & $\rho$ {[}kg/$\text{m}^3${]} & $\nu$ & $\alpha$ & $\beta$ \\ \midrule
    Steel (rail) & 210000 & 7850          & 0.3             & -                & -      \\
    Concrete (sleepers) & 35000 & 2400          & 0.15            & -                & -      \\
    Ballast                       & 150  & 1560          & 0.2             & 0.0439           & 0.0091 \\
    Sand (embankment)             & 80    & 1810          & 0.3             & 8.52             & 0.0004 \\
    Clay (subgrade)               & 25.5 & 1730          & 0.3             & 8.52             & 0.0029 \\
    \bottomrule
    \end{tabularx}
\label{table2}
\end{table}

\subsubsection{Interface and boundary conditions}
The following key interface and boundary conditions were chosen with no separation allowed at any interface:
\begin {itemize}
\item {Rail-sleeper: Rail was connected to the midpoints of the sleeper edges via railpads represented by vertical springs (\textit{k} = $1.2 
 \cdot10^{8}$  N/m) and dashpots (\textit{c} =  $5\cdot10^{4}$  N-s/m) \cite{28}.}
\item Sleeper-ballast, ballast-embankment, embankment-subgrade: surface-to-surface tie constraint (perfect matching of displacements and forces) was used for defining the conditions at these three interfaces.
\item Vertical interface between ballast/ embankment/ subgrade and concrete bridge: a hard contact linear penalty method was used to define the normal behaviour and Coulomb's friction law was adopted to define the tangential behaviour with a frictional coefficient equal to 0.5. (details can be found in \cite{24})
\item The bottom of the subgrade and the bridge were fixed.
\end {itemize}

The simulation was performed in two steps. Firstly a static step was performed to consider the effects of gravity in order to obtain the initial stress state of the model under self-weight. It was followed by a dynamic analysis (full Newton-Raphson method) for 1.75 s with a time step of 0.005 s. The loads that have been considered are: gravity load for static analysis and one moving axle load of 90 kN with velocity of 144 km/hr for dynamic analysis. The load moving in the direction from soft side to stiff side of the transition was simulated using the DLOAD subroutine in ABAQUS \cite{24,29}. Note that the formulation of a design criterion, which is the ultimate objective of this paper, is not qualitatively influenced
by the specific character of the moving load, and hence a simplified loading condition (i.e., a moving constant load) is assumed.

\section{Results and Discussion}
In this section, the results are presented in two parts. Firstly a systematic analysis of the track components is performed by studying the kinematic responses (displacement, velocity and acceleration) of rail, sleepers and ballast, forces in rail pads, and maximum equivalent Von Mises stress at top and at the bottom of sleepers, ballast, embankment and subgrade (for a region of 0.6 m under each sleeper). The kinetic energy (KE) and strain energy (SE) variation in space and time is studied for the track bed layers (ballast, embankment and subgrade). Secondly, the analysis mentioned above is used to present a design criterion for RTZs, and its correlation with permanent deformations is demonstrated by comparing results of the analysis with linear elastic material to the one with a non-linear elastoplastic material behaviour of the ballast.

 \begin{figure}
        \centering  \includegraphics[width=\textwidth,keepaspectratio,trim=0cm 3cm 0cm 0cm,clip]{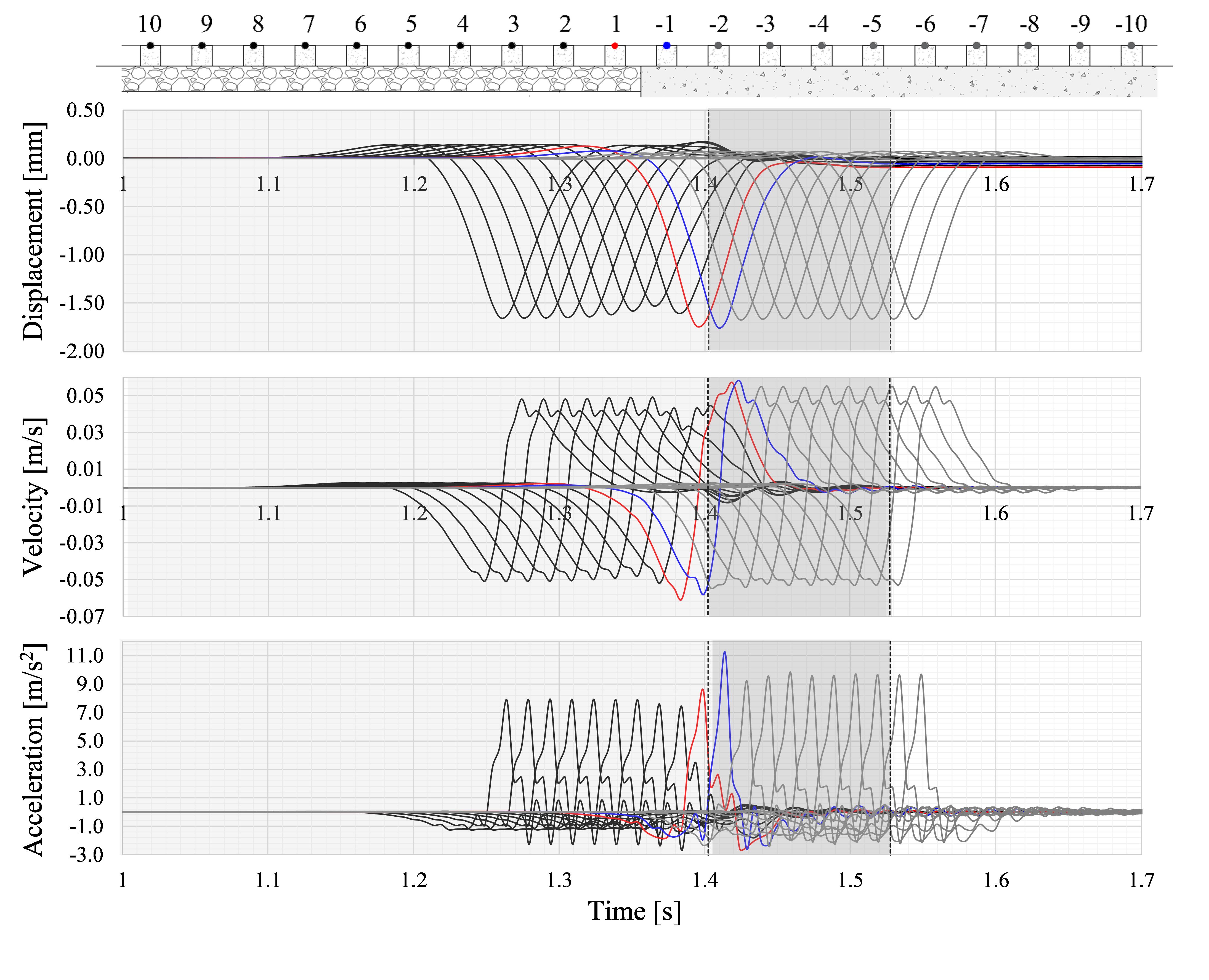}
        \caption{Time history of displacement, velocity and acceleration for 10 control points (on each side of transition interface) of the rail (above sleepers). The dashed lines show the time moments at which the moving load leaves the AZ-I and AZ-II. The red (soft side) and blue lines (stiff side) are the results for the control points above the first sleeper next to transition interface on both sides.}
        \label{fig2}
        \end{figure}

\subsection{Analysis of RTZs}
    \subsubsection{Kinematics of transition zones} \label{firstphase}

        \textit{Rail}: Figure \ref{fig2} shows time history of displacement, velocity and acceleration for 10 control points each in AZ-I and AZ-II. The responses of the rail nodes above the sleepers on both sides right next to the transition show an increase when compared to nodes far from the transition. On one hand the increase in the maximum displacement under the load is only 5.4\%, which can be attributed to the fact that only one axle is being studied. On the other hand, the increase in velocity (20\%) and acceleration (15.5\%) is much larger. Also, some permanent deformation can be seen at locations close to transition enabled by frictional sliding at the transition interface.\\ 
        
\begin{figure}
        \centering  \includegraphics[width=\textwidth,keepaspectratio,trim=0cm 3cm 0cm 0cm,clip]{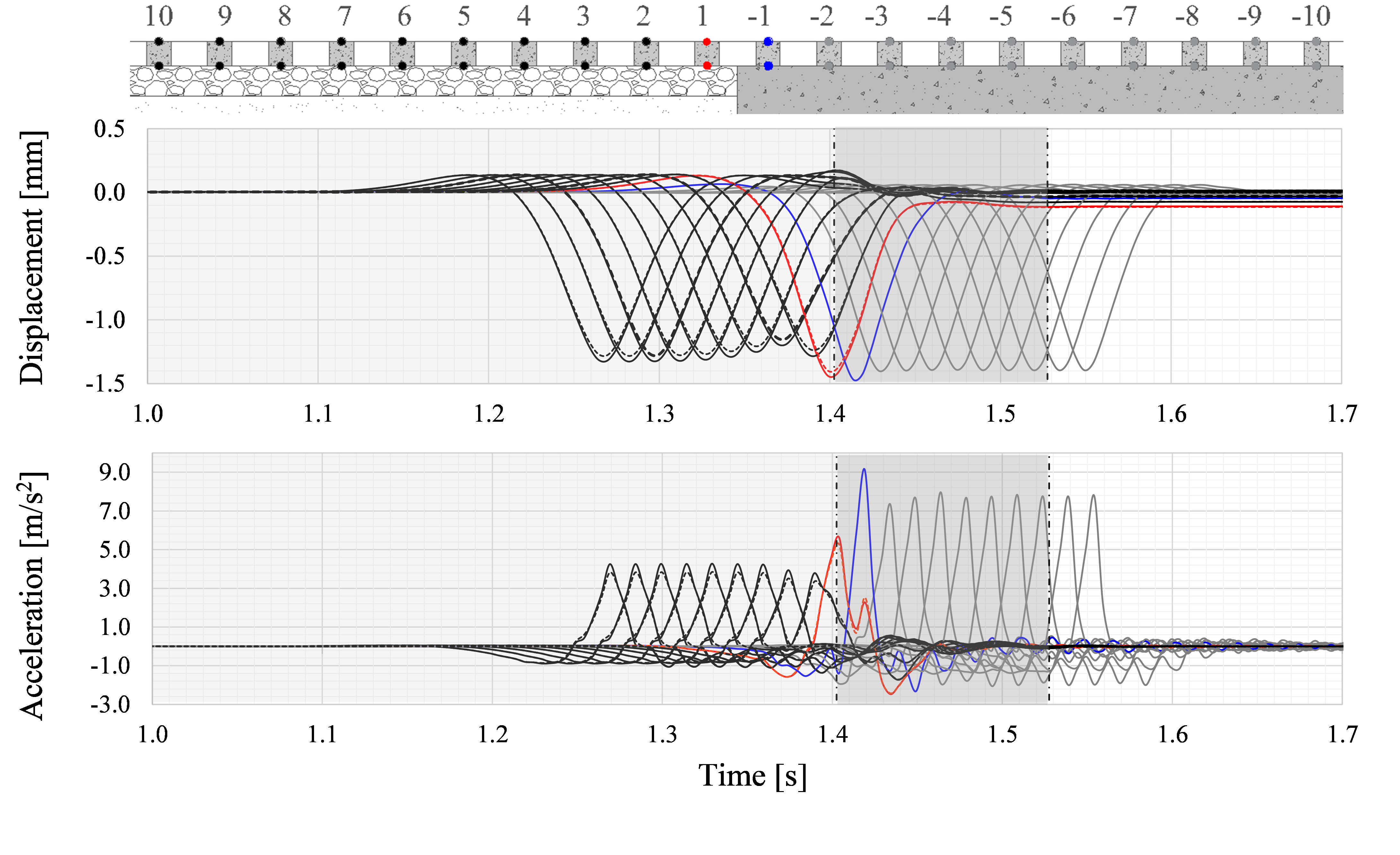}
         \caption{Time history of displacement and acceleration for 10 control points (on each side of transition interface) on top (solid lines) and bottom (dashed lines) of the sleepers. The dash-dotted lines show the time moments at which the moving load leaves the AZ-I and AZ-II. The red (soft side) and blue lines (stiff side) are the results for the first sleeper next to transition interface on both sides.}
        \label{fig3}
        \end{figure}

        \textit{Sleepers}: Figure \ref{fig3} shows time history of displacement and accelerations for 10 control points each (midpoint of sleeper top and bottom) in AZ-I and AZ-II. It can be noticed that 0.114 mm of permanent deformation occurs at sleeper number 1 which can be again attributed to frictional sliding at the transition interface. Although the variation of the kinematic response from top to bottom of the sleepers is negligible, there is a significant increase in the displacements (9.5\%) and accelerations (37.3\%) of the sleepers on left (sleeper 1) and right (sleeper -1) of the transition interface when compared to the sleepers (10, -10, respectively) in far field.\\

\begin{figure}[t]
        \centering        \includegraphics[width=\textwidth,keepaspectratio,trim=0cm 3cm 0cm 0cm,clip]{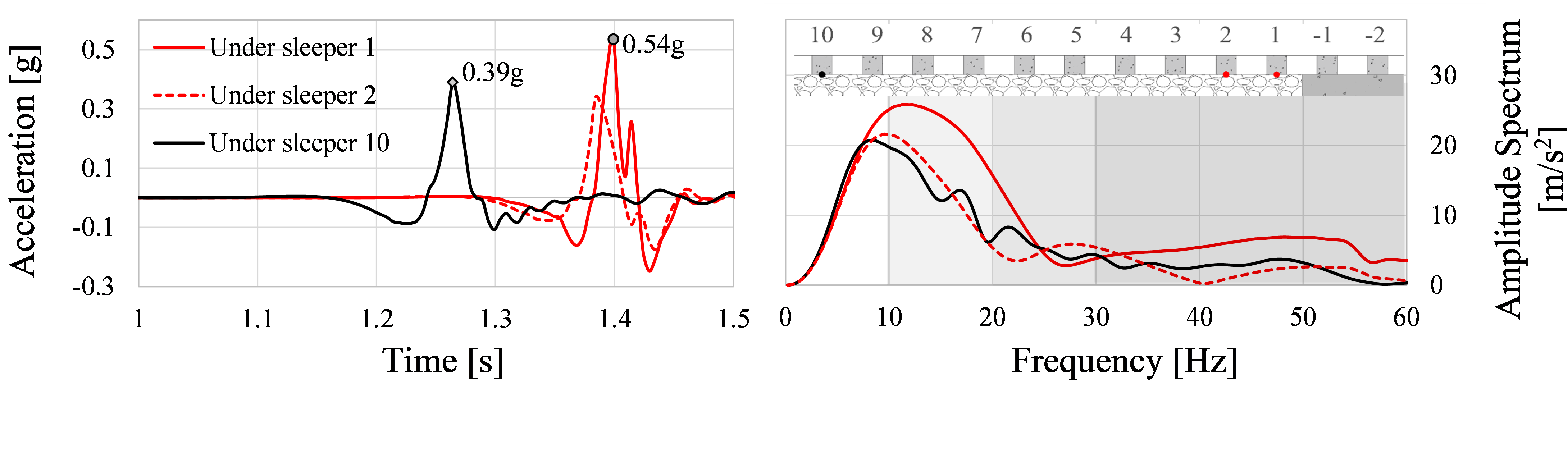}
        \caption{Time history of accelerations(left) and the amplitude spectra of accelerations (right) for two point on top of ballast layer under sleeper 1, 2 and 10.}
        \label{fig4}
        \end{figure}

        \textit{Ballast}: The authors in \cite{30} have associated breakage of ballast particle corners to load frequencies between 10-20 Hz, densification or compaction of ballast without much breakage was associated to load frequencies of 20-30 Hz and all load frequencies above 30 Hz can be associated with splitting of particles. Hence, the amplitude spectra of accelerations for ballast at different locations (under sleepers 1, 2, 10) are analyzed. A significant amplification (38.3\%) is observed in the ballast acceleration under sleeper 1 (0.54g) compared to the ballast acceleration (0.39g) under sleeper 10. It can be clearly inferred from Figure \ref{fig4} that the ballast in the approach zone (under the sleeper 1) is more susceptible to phenomena of corner breakage, compaction and splitting of particles compared to ballast in open tracks, which is in line with what has been observed in reality \cite{31}.

        Clearly, the results in Figure \ref{fig4} show an increase in kinematic quantities at the location of the first sleepers in AZ-I and AZ-II when compared to open track on both the sides. This is in agreement with the current literature \cite{31,32,26} but site measurements show that the first three sleepers (in AZ-I) next to the transition interface are critical in terms of observed damage. This shows that only kinematic studies are not adequate to predict damage in RTZs and for that reason we further investigate forces/ stresses and energies in the following sections.

    \subsubsection{Stresses and forces in transition zones} \label{secondphase}
\begin{figure}
        \centering  \includegraphics[width=\textwidth,keepaspectratio,trim=0cm 3cm 0cm 0cm,clip]{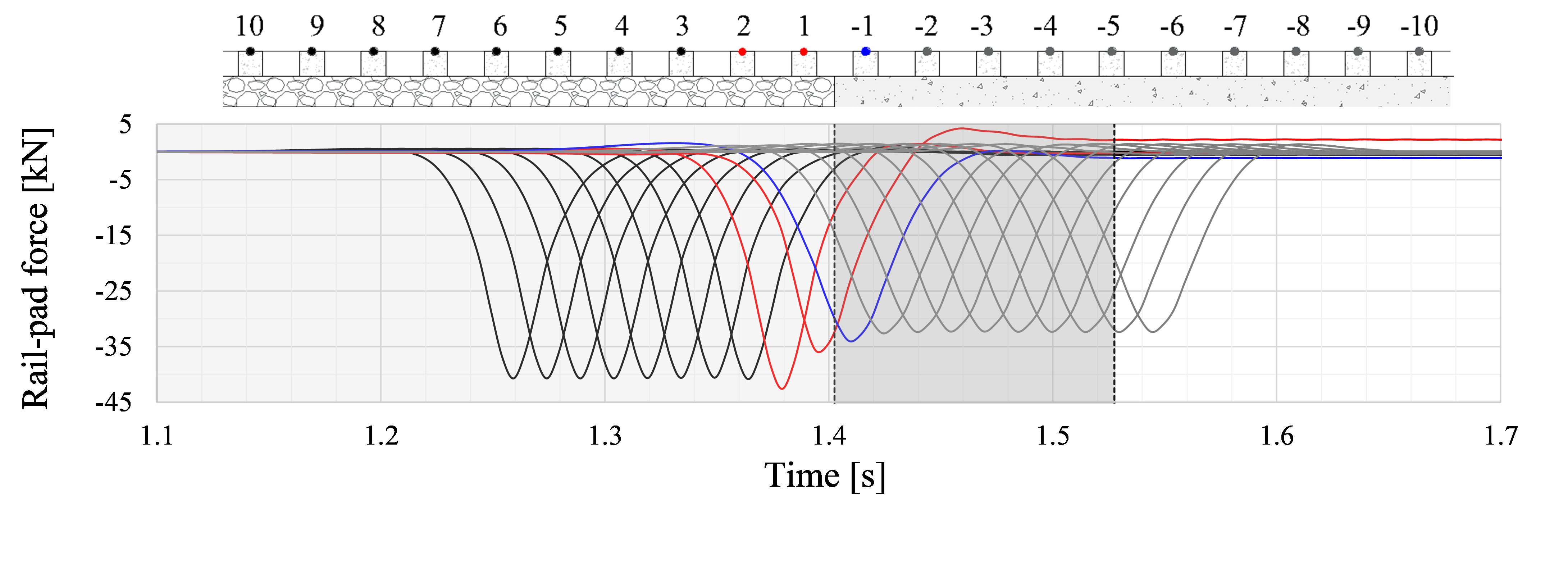}
        
        \caption{Time history of vertical forces in 10 rail-pads next to transition interface on each side. The red lines show the forces in first two rail-pads on the soft side and the blue line shows the force on first sleeper on the stiff side of the transition interface.}
        \label{fig5}
        \end{figure}

        \textit{Rail-sleeper interaction forces}: Figure \ref{fig5} shows a 5\% increase in force (compressive) for the rail-pad connecting sleeper 2 to the rail compared to the force for rail-pad connecting sleeper 10 (in the open track) to the rail. Moreover, the rail-pad connecting sleeper 1 to the rail experiences a tensile force (with respect to the prestress in railpads which is typically present in practice) of approximately 4 kN. In the model, this is due to AZ-I and AZ-II deforming differently, which results in sleeper 1 hanging on the rail. In reality there might be a tensile force or reduction in prestress based on the conditions present at the site.\\
        
\begin{figure}[t]
\centering
\subfigure[]{\includegraphics[width=0.49\textwidth,keepaspectratio,trim=0cm 3cm 35cm 0cm,clip]{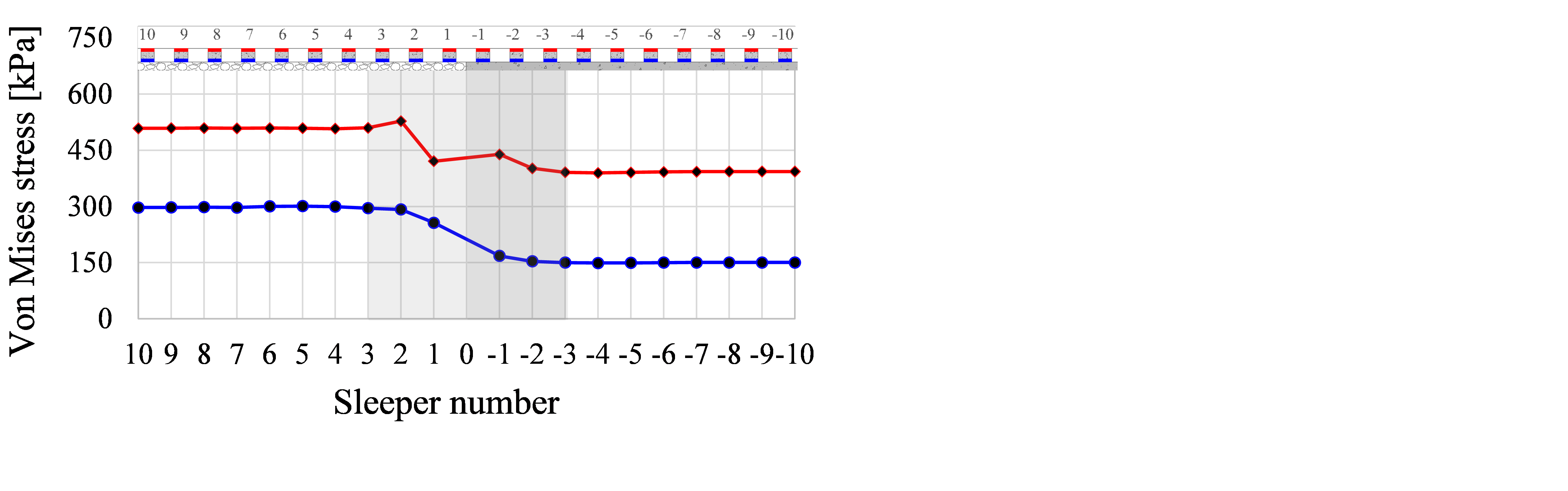}}
\subfigure[]{\includegraphics[width=0.49\textwidth,keepaspectratio,trim=0cm 3cm 35cm 0cm,clip]{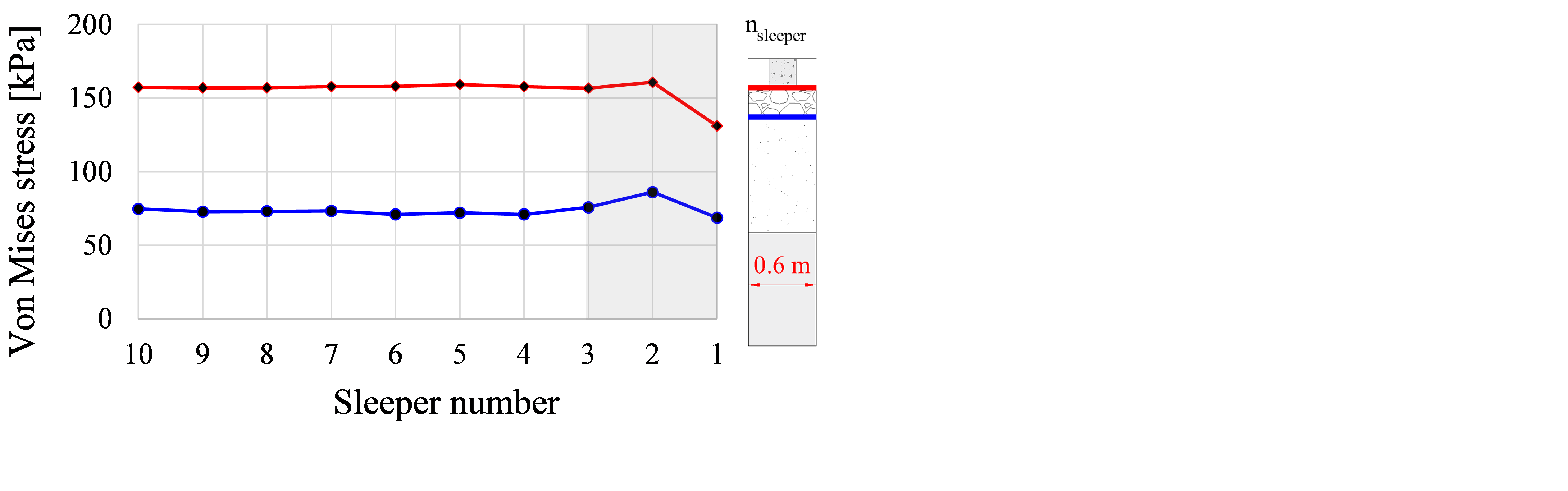}}\\
\subfigure[]{\includegraphics[width=0.49\textwidth,keepaspectratio,trim=0cm 3cm 35cm 0cm,clip]{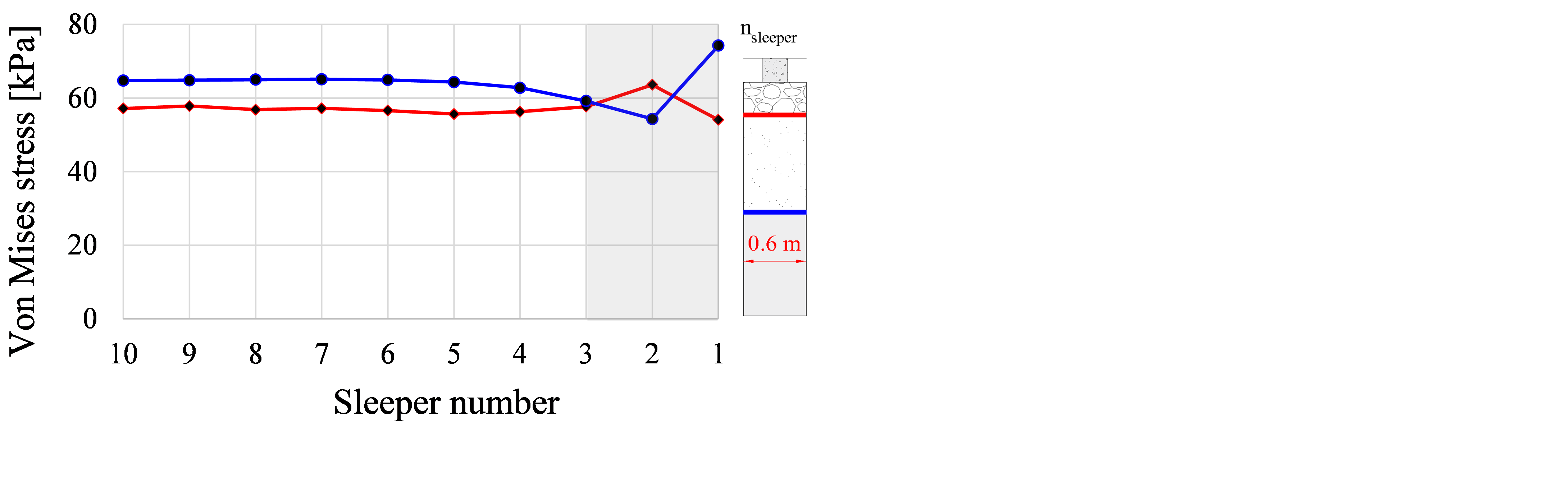}}
\subfigure[]{\includegraphics[width=0.49\textwidth,keepaspectratio,trim=0cm 3cm 35cm 0cm,clip]{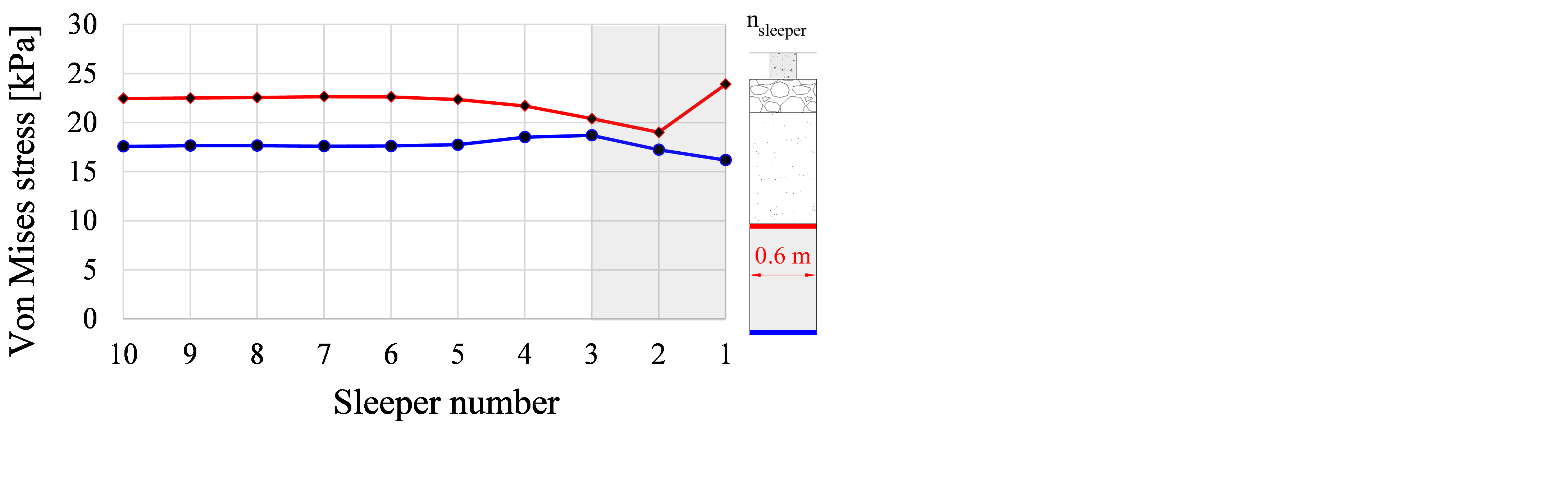}}
\caption{Maximum equivalent Von Mises stress at top (red) and bottom (blue) of (a) sleepers, (b) ballast, (c) embankment, and (d) subgrade for one sleeper bay.}
\label{fig6}
\end{figure}

        \textit{Maximum equivalent Von Mises stress}: In this section the maximum equivalent Von Mises stresses at the top and bottom of each track component (sleepers, ballast, embankment, subgrade) are assessed in terms of their variation along the longitudinal direction (i.e., ‘n’ denotes the sleeper number). According to the Von Mises criterion, yielding of a material begins when the equivalent Von Mises stress exceeds the yield stress of the material. Although the magnitude of the Von Mises stresses are not close to the yield stresses of the materials as the system is analysed only for one cycle and one axle, yet the amplification in the AZ-I compared to OT-I can be clearly seen in Figure \ref{fig6}. This implies that AZ-I will yield earlier than OT-I. The stress amplifications are observed under the first three sleepers.\\

  Figure \ref{fig6} shows that the max. Von Mises stresses under the sleepers 4-10 are the same while there is an amplification under sleepers 1, 2 and 3 (similar to situation on top of the sleepers), which might lead to differential settlement in the long term. Also, a spike in stress can be seen on sleeper 2 and on the ballast under sleeper 2, which is consistent with the amplification in rail-pad force at this particular location (Figure \ref{fig5}). Moreover, the non-uniformity in max. Von Mises stress under sleepers 1, 2 and 3 for embankment and subgrade can be due to a combination of increased force in rail-pad connecting sleeper 2 (Figure \ref{fig5}) to rail and the boundary effects (reflection) at the transition interface.\\

Although  an amplification in max. Von Mises stress in AZ-I compared to OT-I is clearly seen, minimizing these stress values does not necessarily assure that the materials do not fail at lower stresses. Note that the Von-Mises criterion is essentially an energy criterion with the maximum distortional energy density as the critical condition, the distortional energy can be expressed as follows:
\begin{equation}
U_d = \frac{1+\nu}{3E}  \sigma_{VM}^2
\end{equation} 
where $\nu$ is Poisson’s ratio, $E$ is Young’s modulus, $\sigma_{VM}$  is the equivalent Von-Mises stress, and $U_d$  is distortional energy density. The maximum distortion energy theory (Von Mises yield criterion) proposes that the total strain energy can be separated into two components: the volumetric strain energy and the distortional strain energy; the failure occurs when the distortional component reaches the critical value. However, this criterion is most suitable for predicting the yielding conditions of metals and alloys. In RTZs the volumetric strain energy will also play a considerable role in phenomena like breakage, compaction and splitting of ballast particles, eventually resulting in a non-uniform transverse and longitudinal profile of the ballast layer. Although prediction of the type of failure is beyond the scope of this work, it is important to study the total strain energy (as used by authors in \cite{18} to study the failure of soils) in order to assess the degradation in RTZs.\\

As shown in this section, the percentage increase in max. Von Mises stress for the layer of ballast in AZ-I compared to OT-I is 2.57\% and the percentage increase in corresponding distortional energy is 5.2\%. In the following sections, an increase in the total energy will be studied in terms of spatial and temporal variation and the percentage increase will be compared with that of Von Mises stress and distortional energy.

    \subsubsection{Energy variation in transition zones} \label{thirdphase}
\begin{figure}[p]
        \centering
        \subfigure[]{\includegraphics[width=\textwidth,keepaspectratio,trim=3cm 19.8cm 2.5cm 2.5cm,clip]{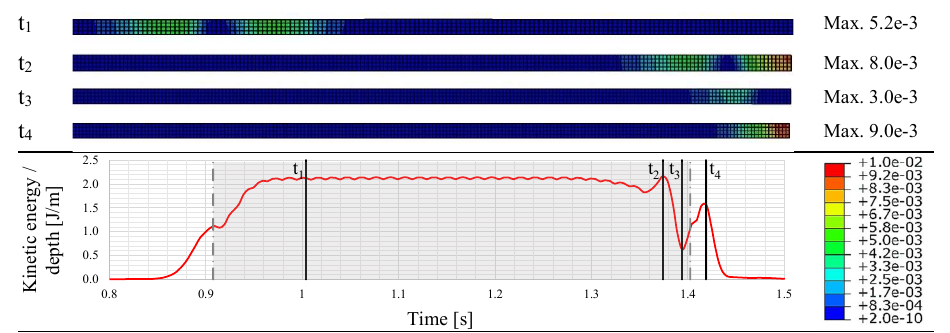}}
        \subfigure[]        {\includegraphics[width=\textwidth,keepaspectratio,trim=3cm 19.7cm 2.5cm 2.5cm,clip]{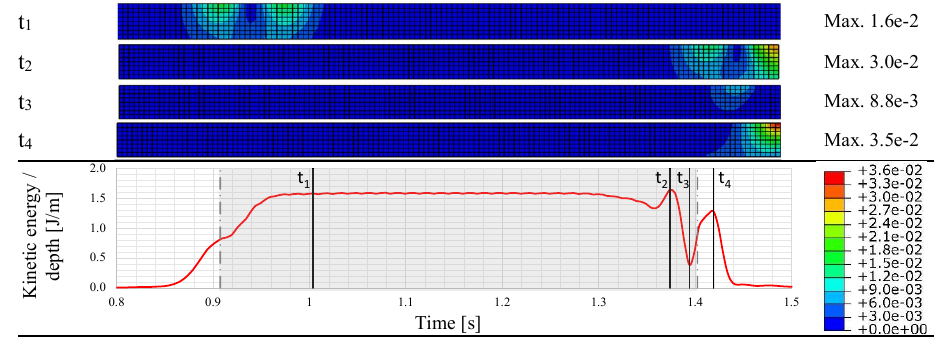}}
        \subfigure[]{\includegraphics[width=\textwidth,keepaspectratio,trim=3cm 19.6cm 2.5cm 2.5cm,clip]{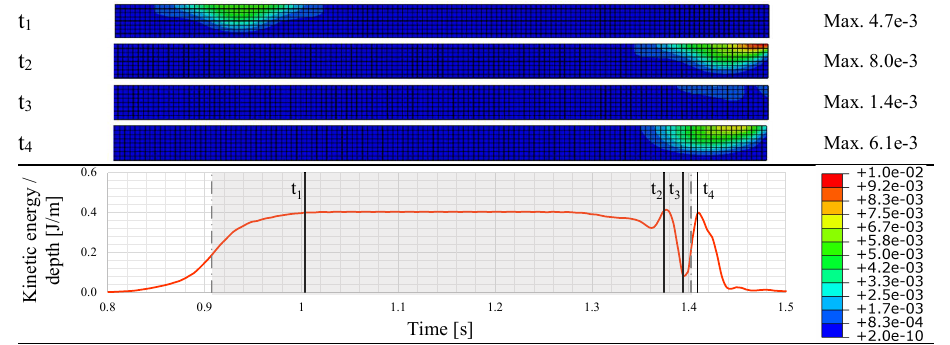}}

        \caption{Spatial distribution of kinetic energy per unit depth in AZ-I in (a) ballast (b) embankment (c) subgrade at time instances $t_{1-3}$. The graphs show the time moments (dashed lines) at which the load enters/exits the AZ-I and the time interval (grey shaded region) for which the load stays in the AZ-I.}
        \label{fig7}
        \end{figure}

    \textit{Kinetic energy}: Figure \ref{fig7} shows the spatial variation of the kinetic energy (plotted for each element with maximum at marked time moments) in ballast (a) embankment (b) and subgrade (c) in the AZ-I at 5 different time instances marked in the figure where a peak ($t_2$, $t_4$) or dip ($t_3$) can be seen in the total kinetic energy and a time instant where the energy is nearly constant ($t_1$). The time instances $t_2$ and $t_4$, when the total kinetic energy has significantly increased, correspond to the load positions above the second sleeper on both sides of the transition interface. It is also interesting to observe that the increase of kinetic energy in the ballast layer is distributed all over the depth (Figure \ref{fig7}a for time instance $t_2$ and $t_4$) and concentrated in region right next to the transition interface. On the contrary this increase is localized for the layers of embankment and soil in the top corner of these layers (also close to the transition interface); see Figure \ref{fig7}b and c. This concludes that the kinetic energy has more significant effect in the ballast layer compared to embankment and subgrade, which could be related to high wear of the ballast particles. Also, it is worth noticing that the elements next to the transition interface experience an increase in kinetic energy even after the load has left.\\
    
\begin{figure}[t]
         \centering  \includegraphics[width=\textwidth,keepaspectratio,trim=0cm 3cm 0cm 0cm,clip]{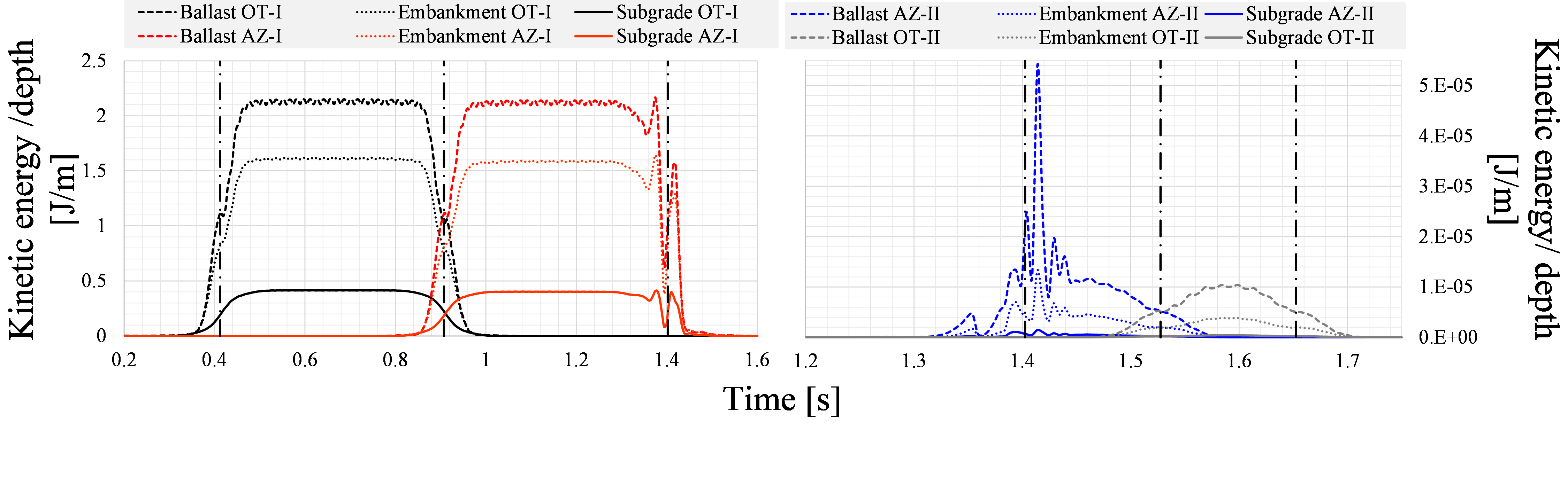}
        \caption{Temporal variation of total kinetic energy per unit depth  in ballast, embankment and subgrade for each zone (OT-I, AZ-I, AZ-II, OT-II)}
        \label{fig8}
        \end{figure}

        Figure \ref{fig8} shows the temporal variation of the kinetic energy per unit depth in each track component (ballast, embankment, subgrade) in all the zones under study (OT-I, AZ-I, AZ-II, OT-II). The dash-dotted lines in the graphs show the time moments at which the load enters and exits the OT-I, AZ-I, AZ-II, OT-II. The total kinetic energy density decreases (approximately 5 times) in magnitude as we investigate deeper in the system whereas the total kinetic energy on soft side is approximately 100 times higher than that on the stiff side.\\

\begin{figure}[p]
        \centering
        \subfigure[]{\includegraphics[width=\textwidth,keepaspectratio,trim=3cm 20.7cm 3cm 2.5cm,clip]{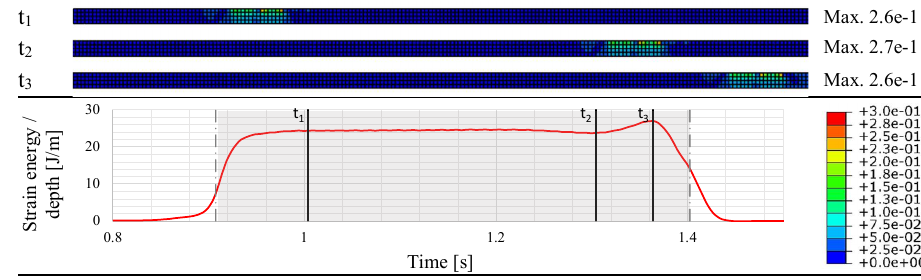}}
        \subfigure[]        {\includegraphics[width=\textwidth,keepaspectratio,trim=3cm 20cm 3cm 2.5cm,clip]{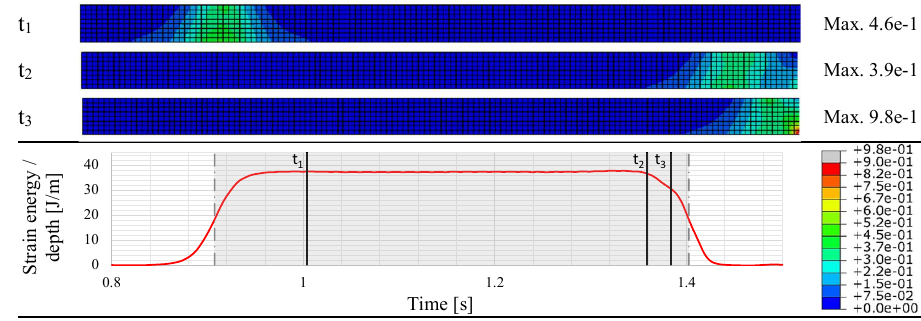}}
        \subfigure[]{\includegraphics[width=\textwidth,keepaspectratio,trim=3cm 19.7cm 3cm 2.5cm,clip]{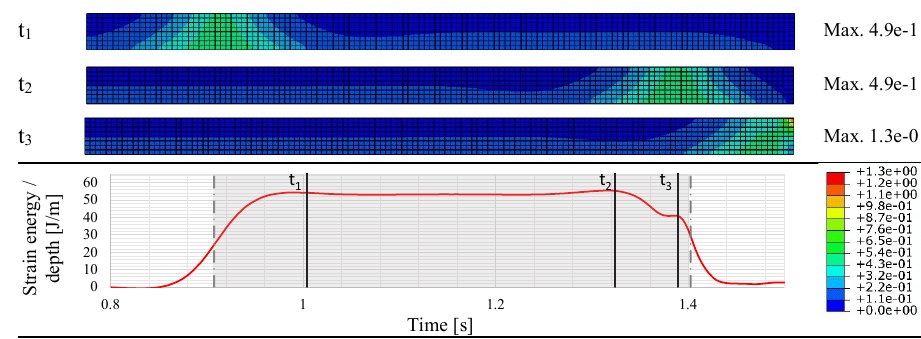}}

        \caption{Spatial distribution of strain energy per unit depth in AZ-I in (a) ballast (b) embankment (c) subgrade at time instances $t_{1-3}$. The graphs show the time moments (dashed lines) at which the load enters/exits the AZ-I and the time interval (grey shaded region) for which the load stays in the AZ-I.}
        \label{fig9}
        \end{figure}

        \textit{Strain energy}: Figure \ref{fig9} shows a spatial variation of strain energy (plotted for each element with maximum at marked time moments) in ballast (a) embankment (b) and subgrade (c) in the AZ-I at 3 different time instances marked in the figure where extremes ($t_2$, $t_3$) can be seen in the total strain energy in ballast layer and a time instant where the strain energy is constant ($t_1$). However, these critical time moments at which the strain energy curve exhibits extremes correspond to different load positions. In the ballast layer, the maximum total strain energy peak occurs at $t_3$, which is the time moment at which the load crosses the third sleeper (soft side). Similarly, a sharp local increase of strain energy can be seen at the bottom of the embankment layer (Figure \ref{fig9}b) at the time instance when the load passes over the second sleeper (soft side). In addition to this, a similar increase can be seen for the subgrade at the time moment when the load crosses the first sleeper (soft side). This shows correspondence to results obtained from both kinematic and stress studies combined together. Moreover, the locations of the strain energy peaks are restricted to regions next to the transition interface (for ballast) and the interfaces between the three track bed layers.\\

\begin{figure}[h!]
        \centering  \includegraphics[width=\textwidth,keepaspectratio,trim=0cm 3cm 0cm 0cm,clip]{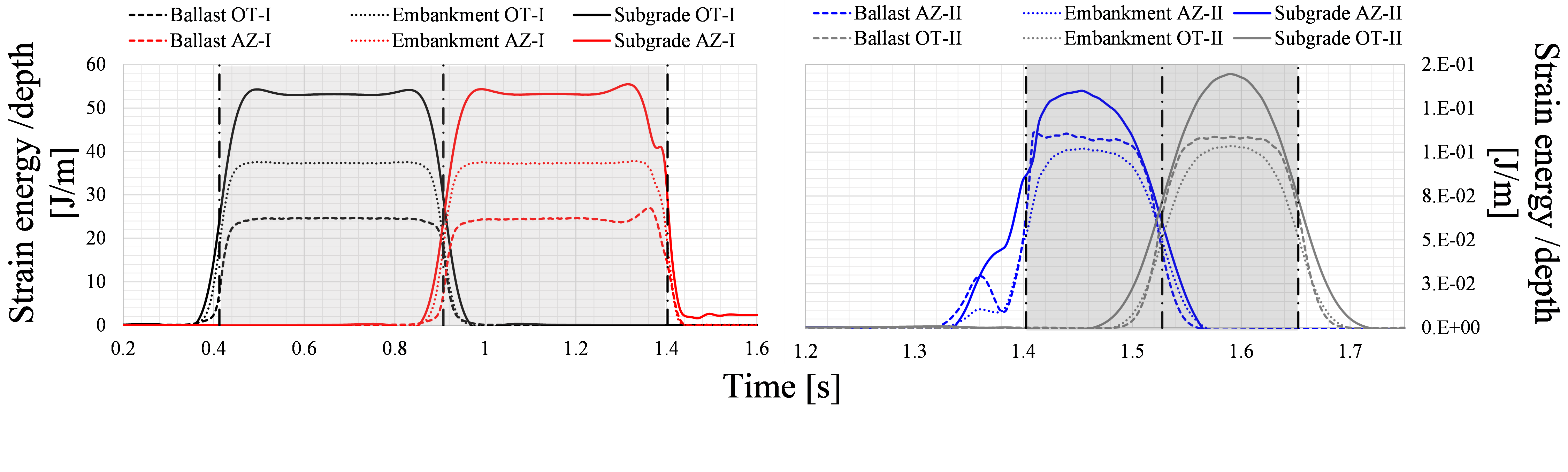}
        \caption{Temporal variation of total strain energy per unit depth in each track component for each zone (OT-I, AZ-I, AZ-II, OT-II)}
        \label{fig10}
        \end{figure}

Figure \ref{fig10} shows the temporal variation of the total strain energy (contribution only due to dynamic components) per unit depth in each track component (ballast, embankment, subgrade) in all the zones under study (OT-I, AZ-I, AZ-II, OT-II). The dash-dotted lines in the graphs show the time moments at which the load enters and exits the OT-I, AZ-I, AZ-II, OT-II. The total strain energy density increases (approximately 2.5 times) in magnitude as we investigate deeper in the system whereas the total strain energy on soft side is approximately 300 times larger than that on the stiff side.\\

\subsection{Design criterion}
A design criterion for RTZs can be proposed based on the comprehensive analysis presented in the first part of this work.\\

In the first phase of analysis in Section \ref{firstphase}, the kinematic study suggested that there is a significant increase in displacements (5.4\% to 9.5\%) and accelerations (15.5\% to 38.3\%) for the track components (rail, sleepers and ballast) in the transition zone compared to open tracks. The location of the peak value of the kinematic responses under study for each of these track components was in the regions around the first sleeper on either side of the transition interface. The critical values of kinematic response occurred at the time moment when the load was passing over the location where the peak response was observed. In summary, a significant amplification in kinematic response was observed for each track component but it was localized to the region around the first sleeper only. Even though a kinematic response provide valuable information regarding dynamic amplifications in transition zones, it cannot be used as a valid design criterion to reduce degradation as site measurements \cite{31,32,26,40} suggest that the degradation in RTZs is observed at locations around the first 3 sleepers next to the transition interface. Therefore, the kinematic response can be a good criterion to evaluate the performance of the upper track components namely rail and sleepers but is insufficient to describe the degradation processes in the lower trackbed layers.\\

\begin{figure}[t]
  \centering  \includegraphics[width=0.75\textwidth,keepaspectratio]{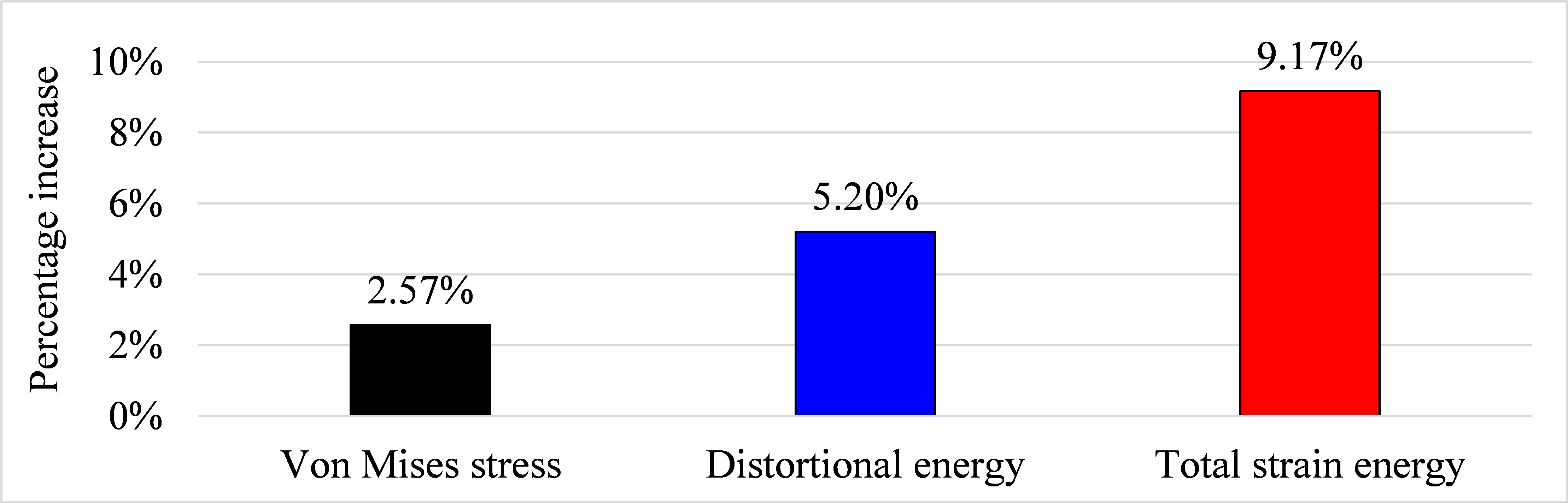}
  \caption{Comparison of percentage increase in Von Mises stress, distortional energy and total strain energy for the ballast layer in AZ-I relative to OT-I}
  \label{fig11}
\end{figure}

The second phase of analysis in Section \ref{secondphase} shows that there is a significant increase in max. equivalent Von Mises stress for each of the track components namely sleepers, ballast, embankment and subgrade in AZ-I compared to the OT-I. The location of the maximum Von Mises stress values for the ballast layer was under the second sleeper next to the transition interface on the soft side of the track. The peak value of the Von Mises stress in embankment layer was localized to the transition interface while in the subgrade layer the peak was in the region around the third sleeper (soft side) next to the transition interface. On one hand, the max. equivalent Von Mises stress shows that the critical locations are similar to those observed on many sites \cite{31,32,26,36,40}. However, on the other hand, based on the discussions above (see Section \ref{secondphase}), the Von Mises criterion takes into account only the distortional component of strain energy and cannot explain all degradation mechanisms that are expected to occur in RTZs. Therefore, to assess or alleviate degradation mechanisms in RTZs, it is more comprehensive to consider the total strain energy, which comprises both distortional and volumetric components. Figure \ref{fig11} summarises the comparison of percentage increase in max. Von Mises stress, distortional energy and total strain energy for the layer of ballast in AZ-I relative to the OT-I. Clearly, the total strain energy is significantly larger than the distortional energy.\\

The third phase of analysis in Section \ref{thirdphase} was focused on the spatial and temporal variation of the mechanical energy. It can be clearly seen from Figure \ref{fig7} and Figure \ref{fig9} that both the total kinetic and strain energies increase by a significant amount for the top track-bed layer (ballast) in AZ-I compared to OT-I. This increase is less prominent in the lower layers (embankment and subgrade). The location of peak values of the energy is in the region around the first to the third sleeper on the soft side of the transition interface. The time moments at which these peak values are observed are when the load passes above sleeper 1 (for ballast), sleeper 2 (for embankment) and sleeper 3 (for subgrade). This shows that energy combines the information obtained from both kinematics and stresses and is therefore a more comprehensive quantity. Now that the magnitude of total kinetic energy is negligible compared to the total strain energy, the authors claim that the degradation of RTZs can be reduced by minimizing the increase in total strain energy in track-bed layers. It is to be noted that the increase in total strain energy for each track-bed layer will depend on various factors such as soil type, degree of compaction of the materials, magnitude and velocity of the load etc. This study shows a significant increase only in the ballast layer due to the choice of model and material parameters, but each layer should be monitored for any increase in total strain energy in the proximity of the transition interface. \\

\begin{figure}[t]
        \centering
        \subfigure{\includegraphics[width=\textwidth,keepaspectratio,trim=0cm 3cm 0cm 0cm,clip]{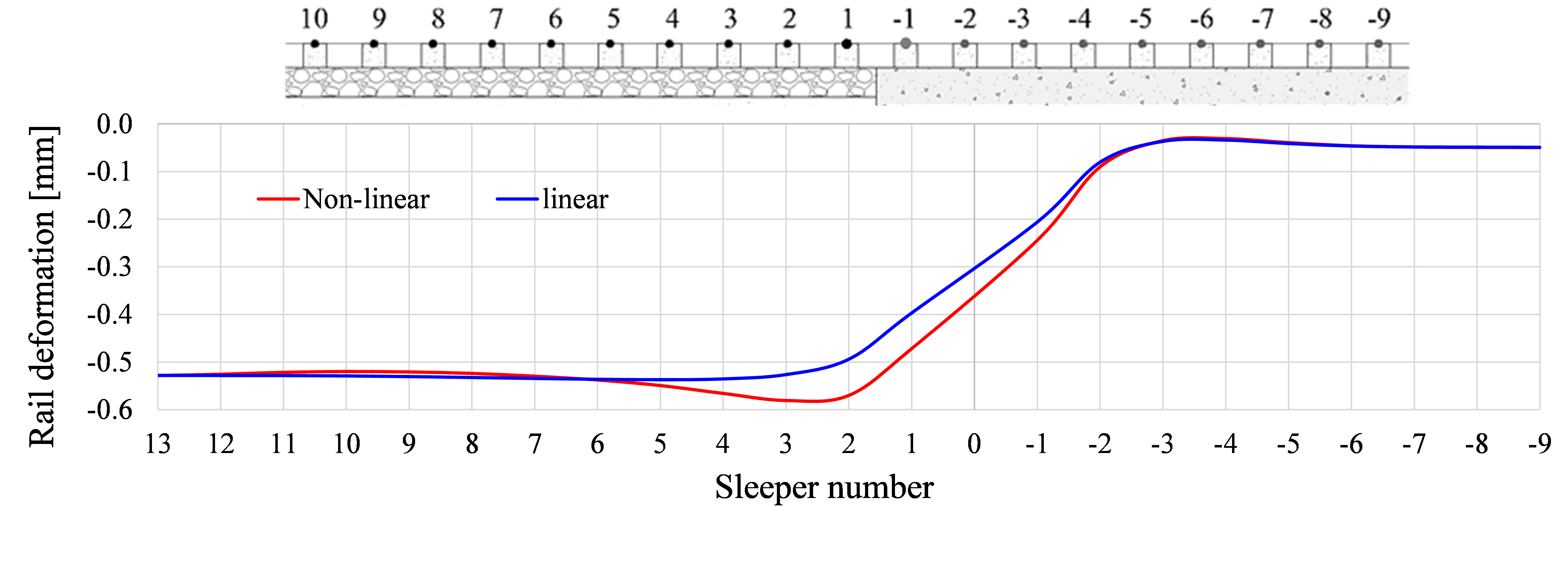}}
        \subfigure       {\includegraphics[width=\textwidth,keepaspectratio,trim=0cm 3cm 0cm 0cm,clip]{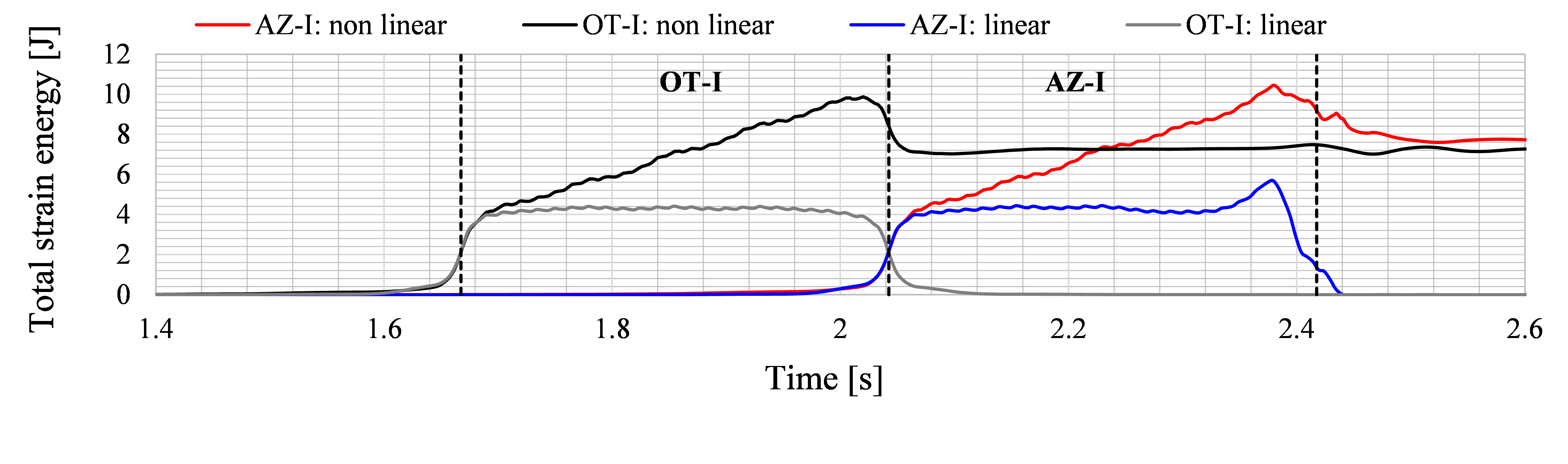}}
 
        \caption{Comparison of (a) rail deformation and (b) total strain energy   obtained from model with linear elastic material behaviour and non-linear elasto-plastic material behaviour of ballast in RTZs for OT-I and AZ-I.}
        \label{fig12}
        \end{figure}

Based on the above discussions and results shown in Section \ref{thirdphase}, it has been claimed that the total strain energy can be seen an indicator of potential damage in RTZs. In reality, a fraction of this total strain energy is recoverable (elastic) and the rest is dissipated through damage. This leads us to the conclusion that in order to minimise the degradation in RTZs, it is evident that there is a need to minimise the total strain energy in each of the track-bed layers. Moreover, a uniform distribution of the total strain energy along the longitudinal direction of track will ensure that there is no non-uniformity in degradation due to localised amplifications in the proximity of a transition interface. The validity of the proposed criterion can be demonstrated by establishing a correlation between permanent irreversible deformation (obtained from model 2) and total strain energy (obtained from model 1). This is done by comparing the results obtained from model 1, which includes linear elastic behaviour of the materials, and model 2, which includes a non-linear elastoplastic behaviour of the material representing the layer of ballast.  The material model \cite{34,35}, the stress-strain relationship and the damage parameters \cite{33} used for simulating the non-linear elastoplastic material (model 2) comparable to ballast are adopted from literature and are used only for demonstration purposes. Hence, the authors do not claim that the absolute values of the quantities shown are fully representative of the railway track materials.\\

Figure \ref{fig12} shows the permanent deformation observed at rail level above each sleeper (a) and the variation of the total strain energy for the layer of ballast (b) in OT-I and AZ-I for both linear elastic (model 1) and non-linear elastoplastic (model 2) material behaviour. It is to be noted that the rail deformation shown in Figure \ref{fig12}b is non-zero at all locations because of the static load due to gravity being active at all time moments. Moreover, the model with linear elastic material shows a very small permanent deformation under sleepers 1 and 2 enabled by frictional sliding at the transition interface. As for model 1, the total strain energy for a finite volume comprises solely recoverable strain energy, while in model 2, it accounts for both recoverable and dissipated strain energy resulting from damage. This explains the increasing trend (Figure \ref{fig12}b) of the total strain energy (in model 2) with the moving load progressing in the given volume. The dashed lines in figure?? represent the time moments at which the load enters and leaves OT-I and AZ-I. The increase in peak value of the total strain energy in AZ-I with respect to OT-I for model 1 is 1.28 J, and for model 2 is 0.57 J. The difference in the values obtained for linear and non-linear models is related to the non-linear material undergoing permanent deformation (Figure \ref{fig12}a) in AZ-I, which is also reflected in the final strain energy difference (0.48 J) between OT-I and AZ-I, as seen in Figure \ref{fig12}b. In conclusion, the findings (Figure \ref{fig12}) indicate that the total strain energy peak observed in the model with linear elastic material behavior is associated with a localized increase (compared to open track) in irreversible deformation in the vicinity of the transition interface (under sleepers 1, 2, 3; see Figure \ref{fig12}a) in the model with non-linear elastoplastic material. This demonstrates the validity of the design criterion proposed in this section.

\section{Conclusions}
A comprehensive study of the behaviour of a railway transition zone (RTZ) was performed. A systematic and detailed analysis of an embankment-bridge transition was conducted where the spatial and temporal distribution of displacements, velocities, accelerations, forces, stresses and mechanical energy were thoroughly studied for the main track components. The investigation in the first part of the paper resulted in mapping the location and extent of the transition effects and formed the basis for formulation of a relevant design criterion that can be used to mitigate the amplified degradation of RTZs. The second part of the paper proposes the total strain energy based on model with linear elastic materials as an indicator to predict degradation. The validity of this proposal was demonstrated and a clear correlation was found between the peak in the total strain energy (in model with linear elastic material behavior) and the localised increase in irreversible deformation (in model with non-linear elastoplastic material behaviour) in the proximity of the transition interface. Therefore, it is concluded that the degradation in railway transition zones can be reduced by minimising the total strain energy in each track-bed layer in the approach zone or to the very least minimising the amplification and non-uniformity of the total strain energy in approach zones relative to the open track. It is claimed that minimising the magnitude of total strain energy will imply lesser permanent deformation and hence reduced degradation, and a uniform distribution of total strain energy along the longitudinal direction of the track will imply uniform degradation.

\section{Acknowledgements}
This research is supported by the Dutch Technology Foundation TTW (Project 15968), a part of the Netherlands Organisation for Scientific Research (NWO), and which is partly funded by the Ministry of Economic Affairs.



 \bibstyle{elsarticle-num} 
 \bibliography{references.bib}





\end{document}